\documentclass[checkin,showpacs,aps,prx,onecolumn]{revtex4-1}

\usepackage{mathrsfs}

\usepackage{color}

\newcommand{\bn}{\begin{enumerate}}
\newcommand{\en}{\end{enumerate}}
\newcommand{\ba}{\begin{eqnarray}}
\newcommand{\ea}{\end{eqnarray}}

\usepackage{graphicx,color}

\newcommand{\be}{\begin{equation}}
\newcommand{\ee}{\end{equation}}

\newcommand{\et}{{\it et al. }}
\newcommand{\ete}{{\it et al.}}

\begin{document}

\newcommand{\clr}{}
\newcommand{\clra}{}











\title{ Spin-orbit torque-mediated spin-wave excitation as an
  alternative paradigm for femtomagnetism }


\author{G. P. Zhang$^*$ and M. Murakami}

 \affiliation{Department of Physics, Indiana State University,
   Terre Haute, IN 47809, USA }

\author{Y. H. Bai}

\affiliation{Office of Information Technology, Indiana State
  University, Terre Haute, IN 47809, USA }

\author{Thomas F. George}

\affiliation{Office of the Chancellor and Departments of Chemistry \&
  Biochemistry and Physics \& Astronomy, University of
  Missouri-St. Louis, St.  Louis, MO 63121, USA }

\author{X. S. Wu}

\affiliation{School of Physics, Nanjing University, Nanjing 210093, China}

\date{\today}

\begin{abstract}
  {Laser-induced femtosecond demagnetization, femtomagnetism, offers a
    potential route to develop faster magnetic storage devices. It is
    generally believed that the traditional spin-wave theory, which is
    developed for thermally driven slow demagnetization, can not
    explain this rapid demagnetization by design.  Here we show that
    this traditional spin-wave theory, once augmented by laser-induced
    spin-orbit torque, provides a highly efficient paradigm for
    demagnetization, by capturing low-energy spin-wave excitation that
    is absent in existing mechanisms.  Our paradigm is different from
    existing ones, but does not exclude them.  Microscopically, we
    find that optical spin-orbit torque generates massive spin waves
    across several hundred lattice sites, collapsing the long-range
    spin-spin correlation within 20 fs.  Our finding does not only
    explain new experiments, but also establishes an alternative
    paradigm for femtomagnetism. It is expected to have far-reaching
    impacts on future research.
}
\end{abstract}




 \maketitle


\section{Introduction}

\begin{figure}
\includegraphics[angle=0,width=0.8\columnwidth]{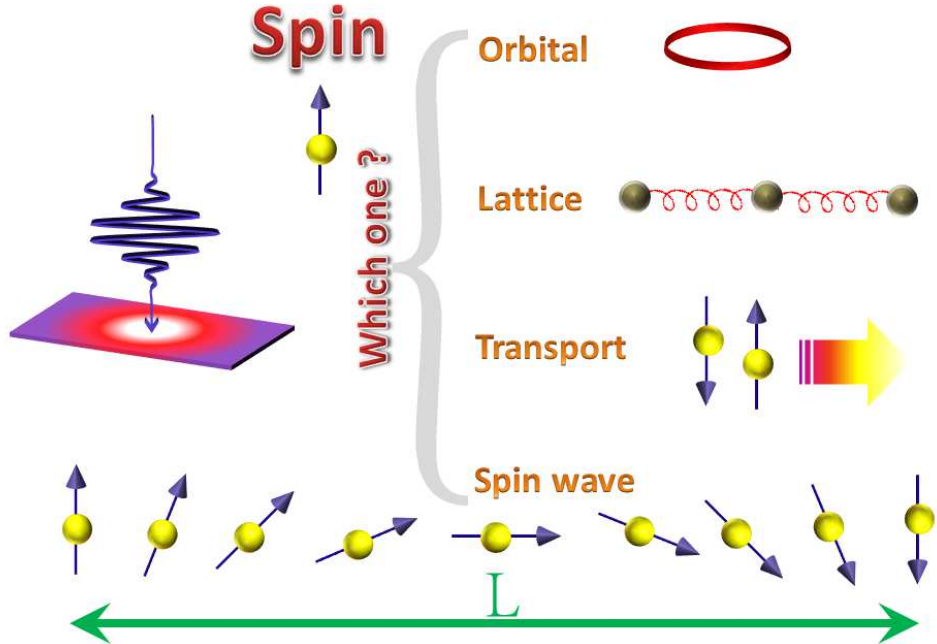}
\caption{Laser-induced ultrafast demagnetization has several possible
  mechanisms from spin-orbit coupling to spin waves. On the left, we
  show a schematic of a typical experiment, where a laser pulse
  impinges on a ferromagnetic thin film, with a finite radius and
  penetration depth. Bottom: {\clr Alternative} paradigm of
  femtomagnetism based on the spin-wave picture, where a spin wave may
  cover many lattice sites. }
\label{fig1}
\end{figure}

Magnetic storage technology is the backbone of information technology
and relies on a fast manipulation of magnetic bits in hard
drives. However, the traditional norm using a faster magnetic field to
flip spins leads to a nondeterministic switching
\cite{stoehr2006}. Twenty years ago, Beaurepaire and coworkers
\cite{eric} discovered that a femtosecond laser pulse could
demagnetize nickel thin films on a much faster time scale. Such a
rapid demagnetization is generic and has been found in various
magnetic systems \cite{ourreview,rasingreview,walowski2016}.  But
despite enormous efforts, its underlying mechanism is still under
intense debate \cite{boeglin2010}.  To this end, at least four
mechanisms have been proposed (see Fig. \ref{fig1}).  {\clr The role
  of spin-orbit coupling in demagnetization was recognized first
 in  \cite{prl00}, and shows up in many systems
  \cite{wietstruk2011,tows2015,shokeen2017,jpcm19}}.  One suggestion
is that demagnetization proceeds by transferring the spin angular
momentum to the orbital angular momentum through spin-orbit coupling
\cite{carpene2008,krauss2009}, or to the phonon subsystem
\cite{stamm2007,baral2014,dornes2019} through Elliot-Yafet scattering
with phonons or impurities \cite{koopmans2010,illg2013}. {\clr
  However, whether the spin transfers angular momentum to phonons is
  unclear {\clra since they only measured the X-ray diffraction signal \cite{dornes2019}}.}
Another suggestion is spin superdiffusion
\cite{battiato2010,eschenlohr2013}.  Each of proposed mechanisms alone
is insufficient to explain the experimental findings
\cite{turgut2013,schellekens2013,moisan2014,turgut2016,jpcm18,vodungbo2016}.  {\clr
  Spin-wave excitation (magnon) was also proposed several times before
  \cite{vankampen2002,melnikov2003,carpene2008,schmidt2010,haag2014,knut2018}.
} It is possible to use spin transfer torque to drive spin dynamics
\cite{razdolski2017}.  Recently Eich and coworkers \cite{eich2017}
reported the band mirroring effect, suggesting spin flipping, and
Tengdin \et \cite{tengdin2018} reported an ultrafast magnetic phase
transition with 20 fs.  More recently Chen and Wang \cite{wang2018}
revealed that spin disorder is crucial to demagnetization. But further
investigations in spin-wave excitation have been scarce
\cite{iacocca2018}. The direct connection between demagnetization and
spin-wave excitation is missing.

Here we aim to establish an alternative paradigm for laser-induced
femtosecond demagnetization that is based on the spin-orbit torque and
rapid spin-wave propagation.  We employ a system of more than 1
million spins and adopt an atomic spin model that includes both the
exchange interaction between spins and spin-orbit coupling, and takes
into account the interaction between the laser pulse and our system
realistically.  {\clr We find that the laser-induced spin-orbit torque
\cite{razdolski2017}} starts a chain of action from spin flipping to
demagnetization, triggers the spin precession, and generates a massive
spin wave.  The spin wavelength is over hundreds of lattice sites, so
the energy cost is extremely low.  A laser pulse of fluence 0.1
mJ/cm$^{2}$ can quench the spin angular momentum of 0.3$\hbar$ by
$-20\%$.  We observe the long-range spin-spin correlation collapsing
within 20 fs, consistent with the experimental observation
\cite{tengdin2018}. Band mirroring \cite{eich2017} is a natural
consequence of spin-wave excitation.  Our finding establishes an {\clr
  alternative} paradigm for laser-induced ultrafast demagnetization
and should have a profound impact in the future.

The paper is arranged as follows. In Sec. II, we develop our concept
of demagnetization through spin-wave excitation. In Sec.  III, we
present our theoretical formalism which includes the model
construction and numerical algorithm. Section III is devoted to the
results. We first investigate how the laser helicity, exchange
interaction, spin-orbit coupling and laser field amplitude affect the
demagnetization, and then demonstrate the collapse of the spin-spin
correlation function. We resolve the spin change in time and space,
and present the images of all the spins in real space. A movie is
provided. The discussion and {\clr alternative} paradigm are presented
in Sec. IV. We conclude this paper in Sec. V.

\section{Spin-wave excitation under laser excitation: Concepts }

 Modern magnetism theory has two competing pictures of magnetic
 excitation. In the itinerant Stoner model, spins are mobile and
 follow the charge, so a band structure description is favored, where
 the spin majority and minority bands shift toward each other, and
 consequently the spin moment is reduced. In the localized Heisenberg
 picture, the spin degree of freedom is decoupled from the charge. The
 module of spin at each lattice site is constant, but the spin
 orientation is not.

Consider a one-dimensional spin chain in the Heisenberg picture. For a
traveling wave and under the linear response approximation, the
transverse components at each site are $s_x^j= u\cos(jka-\omega t)$,
and $s_y^j= u\sin(jka-\omega t)$, where $j$ is the site index, $u$ is
the amplitude of spin wave, $k$ is the spin-wave vector, $a$ is the
lattice constant, $\omega$ is the spin-wave angular frequency, and $t$
is time. The longitudinal component is $s_z^j=s$. At each site $j$,
the spin module is $\sqrt{s^2+u^2}$, which is independent of the site
index. The total spin ${\bf S}$ of the system along the $x$, $y$ and
$z$ axes is $(\sum_js_x^j, \sum_js_y^j, \sum_js_z^j)$. If the
summation is over one wavelength of the spin wave, then ${\bf
  S}=(0,0,Ns)$, where the transverse components all cancel out and $N$
is the total number of lattice sites.

 The Heisenberg picture provides a simple explanation of
 demagnetization.  Consider a simple model with two spins, ${\bf s}_1$
 and ${\bf s}_2$, with modules $|{\bf s}_1|$ and $|{\bf s}_2|$.  The
 total spin is ${\bf S}={\bf s}_1+{\bf s}_2$. To determine the module
 $S$ of ${\bf S}$, we compute ${\bf S}\cdot {\bf S}$ as follows: \be
 S^2={\bf S}\cdot {\bf S} = |{\bf s}_1|^2+|{\bf s}_2|^2+2 {\bf s}_1
 \cdot {\bf s}_2~. \label{twospin} \ee The last term in
 Eq. (\ref{twospin}) depends on the spatial orientation of spins. As
 far as ${\bf s}_1$ and ${\bf s}_2$ misalign with respect to each
 other, the total spin is guaranteed to reduce.  The total spin can be
 reduced to zero if these two spins are antiparallel to each other
 with the same module.  For many spins, the traditional Heisenberg
 Hamiltonian is $ H_0=-J\sum_{ij}{\bf s}_i\cdot {\bf s}_j$, with $J$
 being the exchange coupling.  The spin reorientation is described by
 the magnon excitation and the number of magnons is simulated by a
 Bose-Einstein distribution.  Since the magnetization is proportional
 to $(S-\langle n \rangle)$, where $\langle n \rangle$ is the magnon
 number and $S$ is the total spin in the ground state, as $\langle n
 \rangle$ increases, the magnetization decreases.  How the spin
 reorients does not enter into the theory.

However, under laser excitation we have to explicitly describe how the
spin reorients itself, and in the literature there is no agreement on
how this occurs.  In this study, we propose the spin-orbit coupling
(SOC) as the key driver for spin reorientation because its energy
scale is comparable to the experimental time scale of several hundred
femtoseconds.  It is easy to understand how the spin is reorientated
under SOC.  Suppose that an electron 1 has spin ${\bf s}_1$ with its
neighboring ${\bf s}_2$. The laser field changes the orbital angular
momentum ${\bf l}_1$ due to the dipole selection rule $\Delta l =\pm
1$.  Then ${\bf s}_1$ is subject to an effective field, ${\bf B}_{\rm
  eff}=-J{\bf s}_2+ \lambda {\bf l}_1$.  ${\bf s}_1$ precesses and
finally settles along ${\bf B}_{\rm eff}$. The angle $\alpha$ between
$-J{\bf s}_2$ and ${\bf B}$ is determined by \be \cos(\alpha)={\bf
  B}_{\rm eff}\cdot (-J{\bf s}_2)/(|{\bf B}_{\rm eff}| \left |-J{\bf
  s}_2 \right |)=\frac{s_2^2-\frac{\lambda}{J}{\bf l}_1\cdot{\bf s}_2}
{\sqrt{(s_2^2+\frac{\lambda^2}{J^2}l_1^2-2\frac{\lambda}{J}{\bf
      l}_1\cdot {\bf s}_2)s_2^2}}~. \label{angle} \ee The spin-orbit
coupling is indispensable to spin reorientation. For instance, if
$\lambda=0$, then $\alpha=0^\circ$, so the spin does not reorient. A
nonzero $\alpha$ allows the spin to choose an angle that may be
different from neighboring sites. This introduces a highly
noncollinear spin configuration that can efficiently demagnetize a
sample.

\section{Theoretical formalism}

As can be seen below, in general $\alpha$ is very small.  This
presents a major challenge for numerical calculations. For instance,
for $\alpha=1^\circ$, to construct a spin wave of just one period, one
must include at least 180 lattice sites along one direction to
accommodate such a long spin wave (Fig. \ref{fig1}).  We wonder
whether our prior model that was developed for all-optical spin
switching \cite{jpcm11,epl15,epl16,mplb16,mplb18} could help. This
model is very much similar to the traditional $t-J$ model, where the
spin degree of freedom is taken into account by the Heisenberg
exchange interaction, and the charge degree of freedom is taken into
account by the hopping term between neighboring atoms. In our model,
we add a twist by replacing the tight-binding term by a real-space
kinetic energy and potential energy term, so we can easily include one
extra term, spin-orbit interaction. With these conditions in mind, we
have the Hamiltonian as \cite{jpcm11,epl15,epl16,mplb16,mplb18}, \be
H=\sum_i \left [\frac{{\bf p}_i^2}{2m}+V({\bf r}_i) +\lambda {\bf
    l}_i\cdot {\bf s}_i -e {\bf E}({\bf r}, t) \cdot {\bf r}_i\right
]-\sum_{ij}J{\bf s}_i\cdot {\bf s}_{j}, \label{ham} \ee where the
first term is the kinetic energy operator of the electron, the second
term is the potential energy operator, $\lambda$ is the spin-orbit
coupling, and $ {\bf l}_i$ and $ {\bf s}_i $ are the orbital and spin
angular momenta at site $i$ in the unit of $\hbar$, respectively.  The
potential energy term $V({\bf r}_i)$ takes the harmonic potential. We
also tried other forms of potentials, but we find the harmonic
potential is the best to model our system.  ${\bf E}({\bf r}, t)$ is
the electric field of the laser pulse and has a Gaussian shape in time
and space, as described by \be {\bf E}({\bf r},t)={\bf
  A}(t)\exp[-\frac{(x-x_c)^2+(y-y_c)^2}{R^2}-\frac{z}{d}], \ee where
$x$ and $y$ are the coordinates in the unit of the site number, and
$d$ is the penetration depth of light.  $R$ is the radius of the laser
spot. The field ${\bf A}(t)$ has a Gaussian shape in the time domain,
with the amplitude $A_0$.  We employ left and right circularly
($\sigma^{+}$, $\sigma^-$) or linearly polarized light ($\pi$), with
the polarization plane in the sample plane (see Fig. \ref{fig1}).
{\clr We do not include the relaxation processes since they only play
  an important role on a much longer time scale.  Our model is
  designed to complement the first-principles theory, where a
  calculation with a large number of highly noncollinear spins is not
  possible.  The strength of our model is that we can treat a system
  with lots of spins along different directions.  We will come back to
  this below.  Finally, we emphasize that our model includes the
  electronic excitation from the beginning because of four bracketed
  terms in Eq. (\ref{ham}).  The essence of our model is that the
  charge dynamics of the electrons is described by the harmonic
  potential, while the spin dynamics of the electrons is described by
  the Heisenberg model.  }

We numerically solve the Heisenberg equation of motion \cite{epl16}
for the spin and other observables at site $i$ as follows: \be i\hbar
\frac{d {\bf s}_i}{dt} = [{\bf s}_i,H].  \ee If the number of lattice
sites is $M$, there are $9M$ differential equations, with $3M$ each
for spin, position and velocity.  The time spent also depends on the
laser field amplitude and pulse duration. A typical run needs more
than 48 hours on a single processor computer with 2.1 GHZ, Intel
Xeon(R). We normally output the system-averaged spin at each time
step, but because of the huge volume of data, we only save all the
spins at a few selected time steps.

\section{Results} 

 Our simulation box contains $N_x=N_y=501$ sites along the $x$ and $y$
 axes, with four monolayers ($N_z=4$) along the $z$ axis, with the
 total number of spins over a million. Our system lateral dimension is
 now comparable to the experimental domain size of 140 nm
 \cite{moisan2014,vodungbo2012,bigot2013}. {\clr One big strength of
   our model is that our system size is much larger than any prior
   theoretical simulation.}  Due to our current computer limitation,
 further increase of system size is not possible.  Our laser duration
 is 100 fs, the photon energy is 1.6 eV, and the radius of the laser
 spot is $R=100$ lattice sites, one-fifth of the simulation box.  The
 laser field amplitude is 0.008 $\rm V/\AA$, which can be converted to
 the laser fluence through \cite{jpcm10}, $ F=\sqrt{2\pi} nc\epsilon_0
 A_0^2\tau$, where $n$ is the index of refraction, $c$ is the speed of
 light, the permittivity in vacuum is $\epsilon_0=8.85\times
 10^{-12}~\rm C^2/Nm^2$, and $\tau$ is the laser pulse duration. At
 $\hbar\omega=1.6$ eV, $n=2.45$. For a pulse of 100 fs with
 $A_0=0.008~\rm V/\AA$, $F=0.10435~\rm mJ/cm^2$, which is at the lower
 end of the experimental laser field amplitudes \cite{jpcm10}.  We
 take the bulk Ni's spin angular momentum ($s_z=0.3\hbar$) as an
 example.  We choose $J=0.1$ eV$/\hbar^2$ and $\lambda=0.06$
 eV$/\hbar^2$ for Ni \cite{prl00}.  Other parameters are also
 considered. We list all the parameters used in this paper in Table
 \ref{tab1}.

\begin{table}
\caption{
Parameters used in our simulation. 
}
\begin{tabular}{l|l}
\hline
\hline
System size $(N_x\times N_y \times N_z)$ &  $501\times 501 \times 4$ \\
\hline
Initial velocity {\bf V} ($\rm \AA/fs$) & (0.0, 0.0, 1.0) \\
\hline
Initial spin {\bf s}$_i$ ($\hbar$)  &  $(0, 0, -0.3)$ \\
\hline
Exchange interaction $J$ (eV/$\hbar^2$) & 0.1-1.0 \\
\hline
Spin-orbit coupling $\lambda$ (eV/$\hbar^2$) & 0.01-0.06 \\
\hline
Laser
duration $T$ (fs)  &  100 \\
\hline
 Laser amplitude $A_0$ (V/$\rm \AA$) & 0.002-0.05 \\
\hline
Photon energy $\hbar \omega$ (eV) & 1.6 \\
\hline
 Helicity & $\sigma^{+}, \sigma^-,\pi$\\ 
\hline
 Beam radius $R$ & 100-200\\ 
\hline
Beam center $(x_c,y_c)$ & (250,250) \\
\hline
\end{tabular}
\label{tab1}
\end{table}

\subsection{Dependence of demagnetization on system and laser parameters}

\begin{figure}

\includegraphics[angle=0,width=0.8\columnwidth]{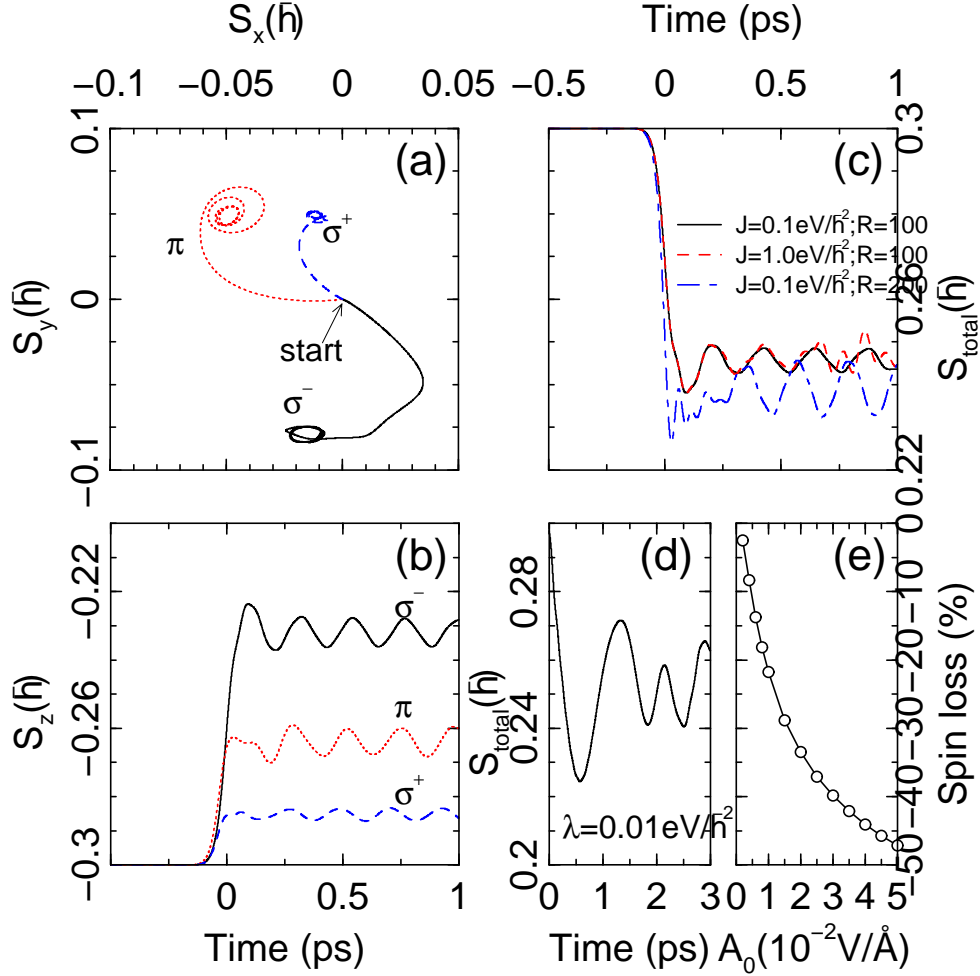}
\caption{(a) Phase diagram of the transverse components of the total
  spin for $\sigma^+$, $\sigma^-$ and $\pi$ light.  The initial point
  is denoted by ``start''.  (b) $z$ component of the spin for
  $\sigma^+$, $\sigma^-$ and $\pi$ light.  $\sigma^-$ has the
  strongest influence because of the spin configuration. (c)
  Dependence of the total spin as a function of time for two exchange
  interactions $J$ and two beam radii $R$. Here a linearly polarized
  pulse ($\pi$) is employed.  (d) Spin reduction with a reduced
  spin-orbit coupling.  (e) Percentage loss of the spin as a function
  of laser field amplitude $A_0$.  }
\label{helicity}
\end{figure}

Once we initialize the spin along the $-z$ axis, $\sigma^+$ and
$\sigma^-$ are no longer equivalent because the spin determines the
quantization axis. Figure \ref{helicity}(a) shows a phase diagram of
the transverse spin components $S_x=\sum_js_x^j$ and $S_y=\sum_js_y^j$
for $\sigma^+$, $\sigma^-$ and $\pi$ light. The word ``start'' denotes
the starting $S_x$ and $S_y$. With arrival of the laser pulse, the
laser helicity has a clear imprint on the phase diagram. For
$\sigma^-$, the spin precesses toward the $+x$ and $-y$ axes before it
settles down at $(-0.016, -0.078)\hbar$, with some oscillations.
Under $\pi$ excitation, the spin swings toward a large $-x$ value and
then settles down at $(-0.047,-0.048)\hbar$, where $S_x$ and $S_y$
have a similar amplitude. $\sigma^+$ has the smallest effect because
it only switches spins from up to down \cite{prb17}. Once the spin
points down, its effect is small. If we compare $\sigma^+$ and
$\sigma^-$, we find their final spins lie opposite to each other. This
strong helicity dependence is consistent with our prior model study
\cite{prb17}.  {\clr It is possible that our model does not include
  the itinerancy of the electrons sufficiently, so the helicity
  dependence is stronger from our model. A weak helicity dependence is
  obtained using the first-principles method \cite{prb17}.} Figure
\ref{helicity}(b) shows the $z$ component of the spin as a function of
time. Regardless of helicity, $S_z$ reduces sharply around 0 ps, and
shifts to a new equilibrium. For each helicity, $S_z$ changes
differently, with the strongest change from $\sigma^-$ (less
negative). The period of the spin oscillation is same.

The helicity is not the only one that affects the spin.  If we
increase the radius of the laser spot to $R=200$, we find the
demagnetization increases (compare the solid and dot-dashed lines in
Fig. \ref{helicity}(c)). To be more objective, Fig. \ref{helicity}(c)
shows the total spin ($|{\bf S}|$).  We find that the increase is much
larger for $\sigma^+$ and $\pi$.  The influence of the exchange
interaction $J$ on demagnetization is particularly interesting, but
its role has been unclear.  We choose two $J$'s, $0.1$ and $1.0$
eV/$\hbar^2$, and plot the results in Fig. \ref{helicity}(c). One sees
that the demagnetization does not strongly depend on the exchange
interaction. Its effect appears only at the later stage. The reason is
simple.  For our initial ferromagnetic spin configuration, the spin
torque due to the exchange interaction is zero and can not flip
spins. The exchange interaction has to wait until the spin flipping
occurs.  Since the spin-orbit torque is the only spin flipping term in
our model, the influence of the exchange interaction is delayed. This
points out an often neglected fact that even though the exchange
interaction is stronger than the spin-orbit coupling, when the
exchange interaction starts to play a role is not determined by the
time scale of the exchange interaction alone. Although spin-orbit
coupling $\lambda$ is small, it dominates the initial spin dynamics.
We can test this idea by reducing $\lambda$ six times to
$\lambda=0.01~{\rm eV}/\hbar^2$.  Figure \ref{helicity}(d) shows the
same total spin as a function of time. We notice that the first spin
minimum appears at 577.7 fs, compared to 90.4 fs at $\lambda=0.06~{\rm
  eV}/\hbar^2$. This is a sixfold increase. Therefore, there is a
direct one-to-one correspondence between the demagnetization time and
the spin-orbit coupling.  However, the amount of the spin reduction
remains similar. To reduce the spin further, we can increase the laser
field amplitude. Figure \ref{helicity}(e) shows a monotonic decrease
of spin up to $-50\%$ with a field amplitude up to $0.05~\rm
V/\AA$. This amount of demagnetization agrees with the experimental
observation \cite{jpcm10,mathias2012}.

\subsection{Spin-spin correlation collapse within 20 fs}

\begin{figure}

\includegraphics[angle=0,width=0.8\columnwidth]{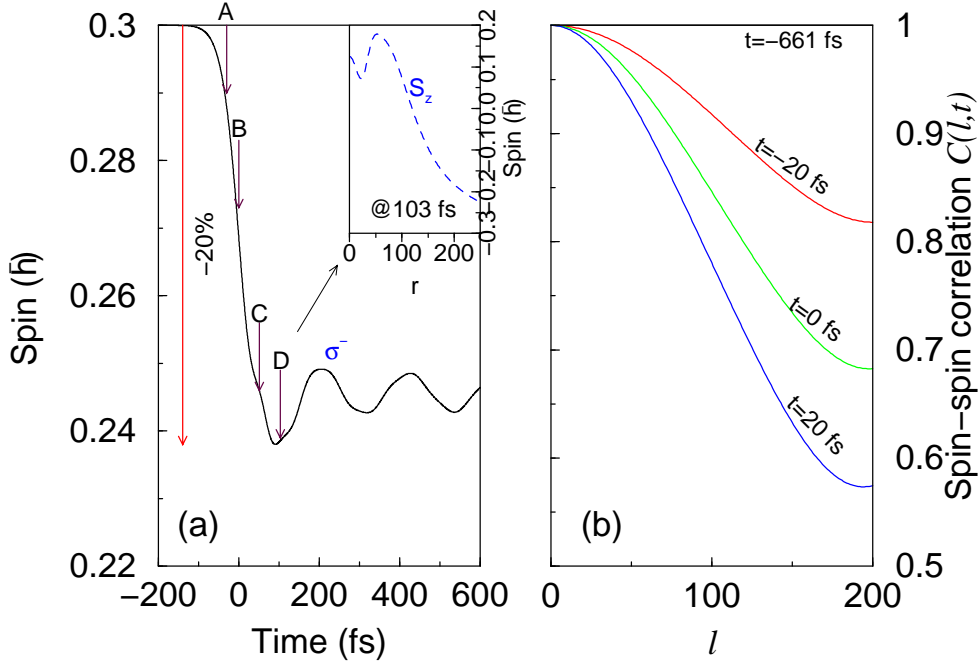}
\caption{(a) Strong demagnetization under a weak laser pulse of
  fluence $0.1~\rm mJ/cm^2$ and duration 100 fs. The spin reduction
  reaches $-20\%$. To be objective, only the magnitude of spin is
  shown.  Four arrows ``A'' to ``D'' denote four times, whose spins
  are spatially resolved in Fig. \ref{image}.  The top inset shows the
  spatially integrated spin at ``D'' (103 fs) as a function of radius
  $r$.  The spins are highly noncollinear, and close to the excitation
  center, the spin tilts toward the $+z$ axis. {\clr (b) The spin-spin
    correlation function $C(l,t)$ at five different times $t$,
    collapsing within 20 fs, as consistent with the experimental
    findings \cite{tengdin2018}.  The correlation carries different
    information from demagnetization. $l$ is the distance between two
    spins. In the ferromagnetic phase, the materials have a long-range
    order}. }
\label{correlation}
\end{figure}

So far, only the phenomenological spin diffusion model
\cite{battiato2010} can reproduce a comparable experimental spin
reduction.  Nearly all the first-principles calculations report a very
small spin moment change \cite{jap08,krieger2015}.  The above huge
spin reduction, without a phenomenological treatment of laser
excitation, is encouraging.  This motivates us to thoroughly
investigate what leads to strong demagnetization. We use
$A_0=0.008~{\rm V/\AA}$ as an example. Figure \ref{correlation}(a)
represents our key result that a weak laser pulse induces $-20$\% spin
reduction within 200 fs. {\clr The spin oscillation is very similar to
  the oscillation in the odd second harmonic signal in Gd measured by
  Melnikov \et \cite{melnikov2003}, but they attributed it to the
  surface phonon vibration. Nevertheless, the fact that our frequency
  of 4.4 THz is comparable to theirs of 2.9 THz shows that their
  magnon wave packets are indeed coupled to the phonons.}
 If we examine our results closer, we find that the spin percentage
 loss is not uniform across the slab.  The inset of
 Fig. \ref{correlation}(a) shows an example at 103 fs (D in
 Fig. \ref{correlation}(a)) of spatially integrated spins with respect
 to the center of excitation, $S_z(r)=\sum^r_{i=1} {s}_{z}^i$,
 where $r$ is the radius of the measurement.  We notice that the spins
 ($S_z$) close to the excitation center reverse their directions to the
 $+z$ axis. Around $r=100$, $S_z$ is close to zero; and away from the
 center $S_z$ remains unchanged since there is no strong
 excitation. 

The above spatial information is useful to understand the phase
transition.  We compute the spin-spin correlation function $C(l,t)$
\cite{iacocca2018}, \be C(l,t)=\sum_{i:x,y} \frac{{\bf s}_i(t)\cdot
  {\bf s}_{i\pm l}(t)}{N_{x,y} |{\bf s}|^2}~, \ee where the summation
is over both the $x$ and $y$ axes, $l$ measures the range of spin
correlation across lattice sites, and $N_{x,y}$ is the normalization
constant. It is expected that the ferromagnetic thin film has a
long-range ordering in the beginning.  {\clr We note in passing that
  this correlation function carries different information from
  demagnetization}.  Figure \ref{correlation}(b) illustrates that the
correlation function is 1.0 long before laser excitation and around
-20 fs reduces to 0.81. At $t=20$ fs, the correlation function drops
precipitously to 0.57 at $l=200$. Collapsing of the spin-spin
correlation is generic, regardless of laser helicity.  When we use a
$\pi$ pulse, the correlation function $C(l,t)$ similarly collapses
within 20 fs.  This nicely explains why experimentally Tengdin \et
\cite{tengdin2018} reported a magnetic phase transition within 20 fs.

\subsection{Space- and time-resolved demagnetization}

\begin{figure}

\includegraphics[angle=0,width=0.45\columnwidth]{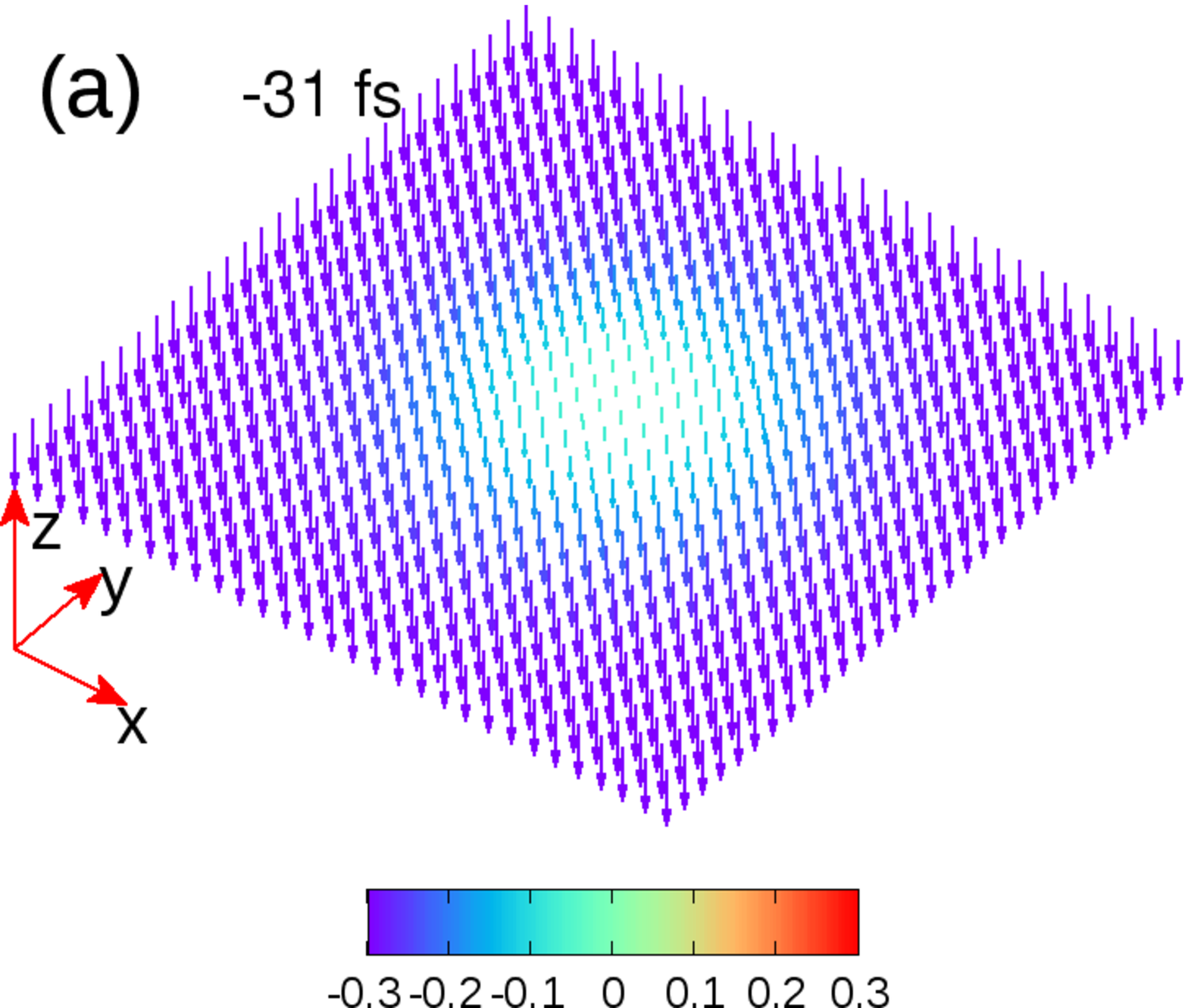}
\includegraphics[angle=0,width=0.45\columnwidth]{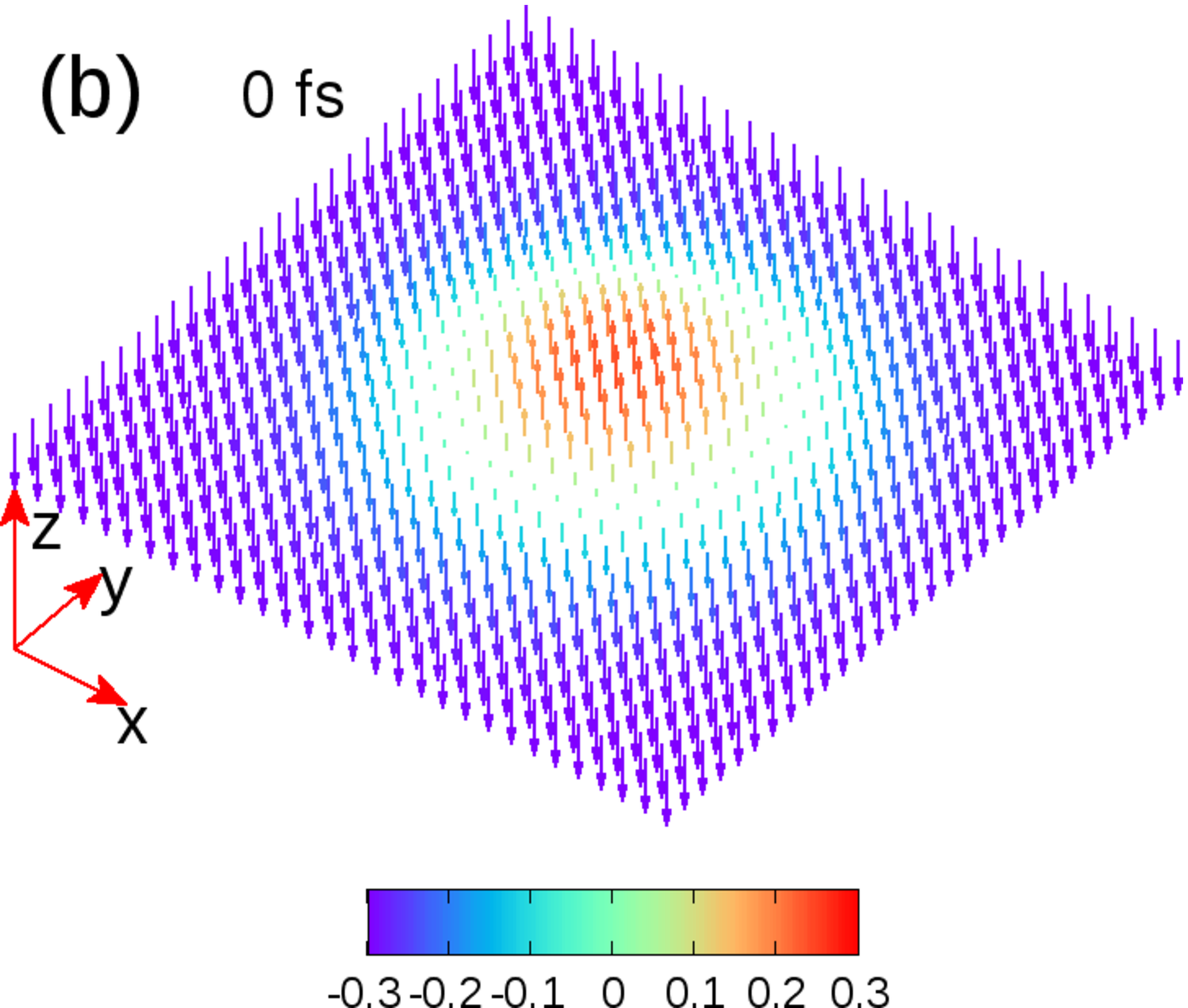}
\includegraphics[angle=0,width=0.45\columnwidth]{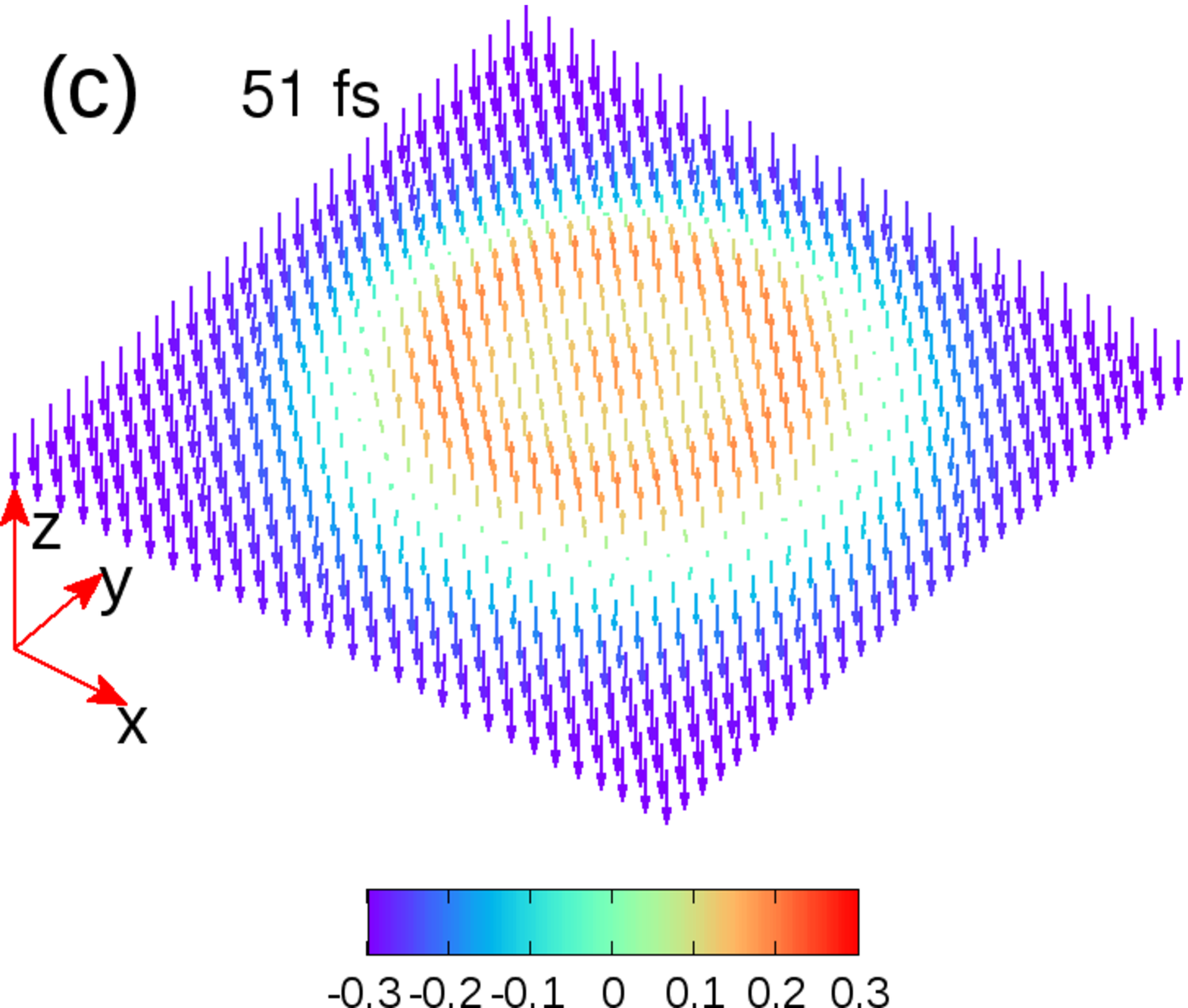}
\includegraphics[angle=0,width=0.45\columnwidth]{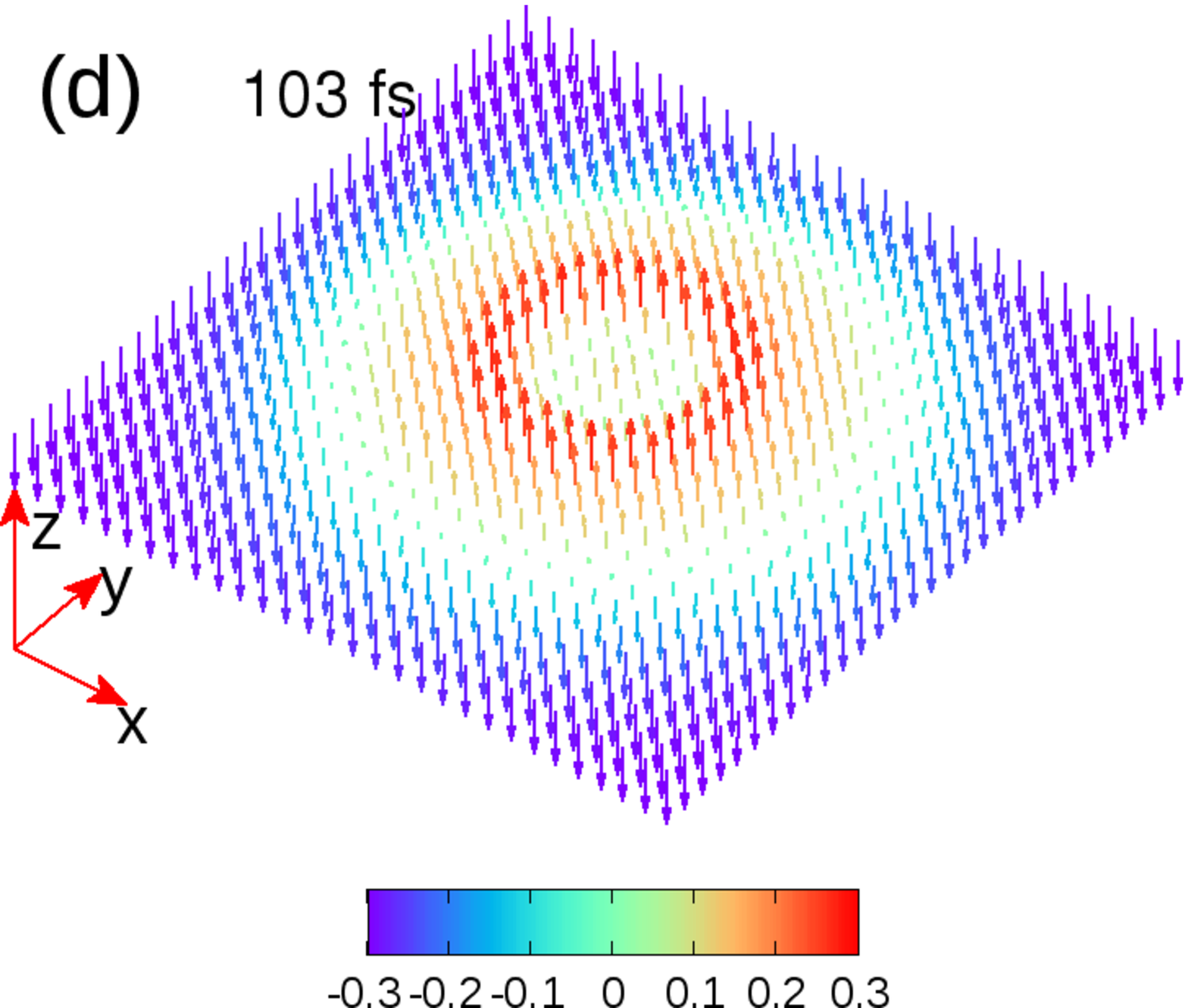}

\caption{Snapshot of spins at (a) $-31$ fs, (b) 0 fs,
  (c) 51 fs and (d) 103 fs.  (a) and (b) show that the demagnetization
  starts with spin flipping at the center.  In (c), spins are highly
  noncollinear and form the spin domains spreading outwards. A spin
  domain wall is built around the center. Spins close to the
  excitation center are reversed. In (d), two layers of spin waves
  appear.  The color bar goes from the violet ($-0.3\hbar$) to the red
  ($0.3\hbar$).  We only show one out of every ten spins within a
  region of $300\times 300$ to reduce the huge volume of data.  
``Multimedia View''.
}
%
\label{image}

\end{figure}

To reveal further insight into demagnetization, it becomes necessary
to resolve the spin change in real space.  We choose four times,
labeled by ``A'' through ``D'' in Fig. \ref{correlation}(a).  Figure
\ref{image} (``Multimedia view'') shows a three-dimensional spin image
snapshot for the entire first layer at these four times of (a) -31 fs,
(b) 0 fs, (c) 51 fs and (d) 103 fs.  The Mutlomedia view has a movie
of a full-time-dependent spin evolution.  To reduce the huge volume of
data to a manageable level, in the figure we only retain one out of
every 10 spins and zoom in the central portion of spins from lattice
site 100 to 400 along the $x$ and $y$ axes. Spins in the vicinity of
our simulation box are not affected and not shown. Figure
\ref{image}(a) shows that at $-31$ fs, the spins tilt away from the
initial $-z$ axis to the $xy$ plane, and the main change is at the
center of the laser pulse, where the field amplitude is the strongest.
Around 0 fs when the laser pulse peaks, spins in a small central
region of 50 lattice sites reverse to the $+z$ axis (see
\ref{image}(b)), and they collectively develop into a spin wave.  But
different from the traditional magnon excitation where the spin
tilting is small, here the spins are highly noncollinear and can be
completely flipped, so the majority (minority) spin becomes the
minority spin (majority), leading to the band mirroring effect
observed experimentally \cite{eich2017}. The spin wave has a
wavelength of 120 sites, or over 40 nm for fcc Ni, which is much
bigger than the unit cell used in prior studies
\cite{krieger2015,krieger2017}.

\begin{table}
\caption{Spin propagation speed $v_p$. The speed of
  \cite{granitzka2019} changes from 0.017 nm/fs to 0.033 nm/fs and
  then back to 0.0148 nm/fs. The table lists the fastest
  speed. Experiments were carried out under different conditions for
  different samples.  The x-ray scattering data consistently show a
  slower speed than that from the holography data. Additional
  near-field measurement is necessary.  XMCD: x-ray magnetic circular
  dichroism.  }

\begin{tabular}{lccc}
\hline
\hline
Sample & $v_p$ (nm/fs)& Method  & Reference \\
\hline
CoPt multilayer & 0.012-0.020  & XMCD+scattering & \cite{pfau2012}\\
\hline
FePt (grain) &0.033  & XMCD+scattering & \cite{granitzka2019} \\
\hline
CoPd & 0.2 & XMCD+Holography
& \cite{vonkorff2014} \\
\hline
 Theory & 0.3-0.4 && This work \\
\hline
\hline
\end{tabular}
\label{tab2}
\end{table}

Figure \ref{image}(c) shows that at 51 fs, the spin wave propagates
outward, and its characteristic time is determined by the exchange
interaction $J$.  For our $J=0.1$ eV$/\hbar^2$, this approximately
corresponds to 40 fs.  The spread of the spin wave is consistent with
those reported experimentally by von Korff Schmising \et
\cite{vonkorff2014}, but they attributed it to spin-polarized electron
transport. We estimate the speed of spin-wave propagation at 1.2
lattice site per fs. Using the fcc Ni or bcc Fe lattice constant, we
have a speed of 0.3 to 0.4 nm/fs, which is surprisingly close to the
experimental value of 0.2 nm/fs \cite{vonkorff2014}, given many
differences between experiment and theory. Therefore, we believe that
their results are more consistent with the spin-wave propagation than
with transport.  If we accept the spin-wave picture, we now can
understand why the demagnetization and spin-wave propagation appear in
tandem. In the spin-wave picture, the spin wave is a precursor of
demagnetization.  In agreement with the experiment by Pfau \et
\cite{pfau2012}, our nanoscale spin walls are established during the
demagnetization, which further explains why their spin correlation
length can expand by 2.8 nm within 0.5 ps, since the spin wall is a
consequence of spin waves and is not a regular magnetic domain wall
which is much slower to establish.  Spin propagation speed $v_p$ is
material dependent.  The expansion of the width $d\xi$ can be computed
from the momentum transfer $dq$, $d\xi=-(2\pi/q^2)dq$. From Pfau's
experiment \cite{pfau2012}, $dq=0.04q_{\rm peak}$, where $q_{\rm
  peak}=42 {\mu \rm m}$, so we get $d\xi=5.98$ nm. Their time has a
large uncertainty of 200 fs. Their $v_p$ is estimated between 0.012
and 0.020 nm/fs.  A faster speed than that in CoPt was found in FePt
grains \cite{granitzka2019}. Table \ref{tab2} summarizes the
experimental results currently available, in comparison with our
theory.  Figure \ref{image}(d) shows that at 103 fs, the perimeter of
the spin wall is defined, and spin waves can recoil from the wall. A
hole is developed in the center. The further spread of the spin wave
ceases from now on. We monitor the entire spin precession up to 900 fs
and find no change.


\section{Discussion: An alternative paradigm}


\begin{figure}
\includegraphics[angle=0,width=0.85\columnwidth]{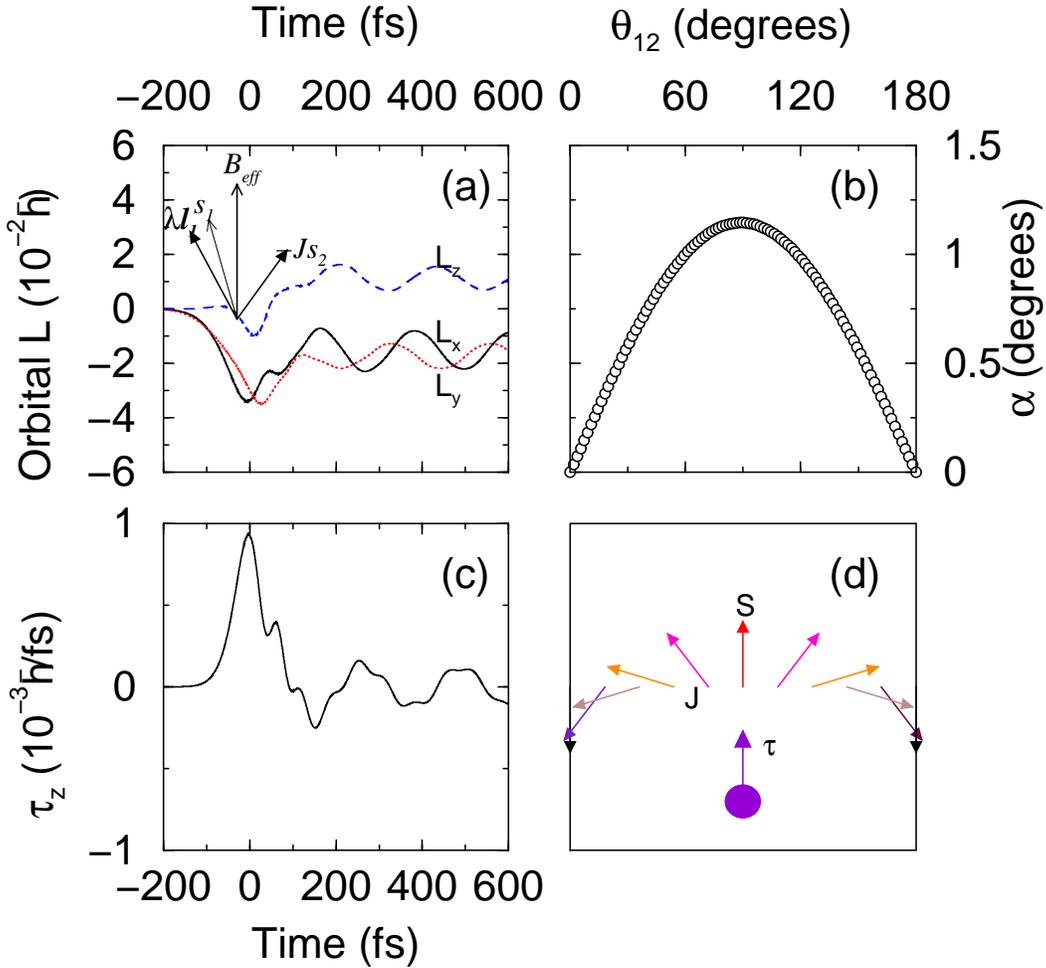}
\caption{(a) Time-evolution of system-averaged orbital angular
  momenta, $L_x$ (solid line), $L_y$ (dotted line) and $L_z$
  (dashed). The overall change is small, within $0.01\hbar$.  Inset:
  Competition between the spin-orbit coupling and the exchange
  interaction.  (b) Spin tilting angle $\alpha$ as a function of angle
  between the orbital angular momentum ${\bf l}_1$ at site 1 and spin
  ${\bf s}_2$ at site 2. A maximum is found at 90$^\circ$.  (c)
  Time-dependent spin-orbit torque. A large positive $\tau_z$ around 0
  fs flips the spins.  (d) Schematic of our {\clr alternative}
  paradigm. The laser induces a strong spin-orbit torque $\tau$ on
  spin $S$. Through the exchange interaction $J$, the spin wave
  propagates. The spatial misalignment leads to a huge spin
  reduction. }
\label{fig4}
\end{figure}

Thermal demagnetization proceeds through a collective low-energy spin
excitation driven by a thermal field. Laser-induced demagnetization is
more complex.  If the demagnetization proceeds through angular
momentum transfer from the spin subsystem to the orbital subsystem,
one should expect a surge of the orbital angular momentum during
demagnetization, though it is not necessary.  We can directly check
this in our model.  Figure \ref{fig4}(a) shows the system-averaged
orbital angular momentum is extremely small, on the order of 0.01
$\hbar$ for all three components, $L_x$, $L_y$ and $L_z$, in agreement
with the experiment \cite{lopez2012}.  This caps the maximum spin
reduction through angular momentum transfer by no more than 3\%.

%
%


How does such a small orbital momentum induce strong demagnetization?
The inset in Fig. \ref{fig4}(a) illustrates that each spin is subject
to at least two competing interactions: spin-orbit coupling and
exchange interaction between neighboring spins. The spin tilting angle
$\alpha$ between a spin ${\bf s}_1$ and its neighboring spin ${\bf
  s}_2$ is determined by Eq. (\ref{angle}).  Since $J$, $\lambda$ and
${\bf s}$ are fixed, the only free parameter is the angle
$\theta_{12}$ between ${\bf l}_1$ and ${\bf s}_2$.  The laser field
enters through ${\bf l}_i$.  It is important to realize that the
direction of ${\bf l}_1$ is controlled by the laser pulse, not by
spin, so $\theta_{12}$ can take any value between 0$^\circ$ and
180$^\circ$. We plot $\alpha$ as a function of $\theta_{12}$.  Figure
\ref{fig4}(b) shows $\alpha$ has a maximum of about 1 degree at
$\theta_{12}=90^\circ$. This means that even assuming $\alpha$ remains
constant, one needs 180 lattice sites to get a spin configuration
completely reversed at the last lattice site with respect to the first
site. This number matches our domain width of 240 sites. For each
tilting, there is an energy cost due to the exchange interaction
$-J{\bf s}_1\cdot {\bf s}_2$.  The exchange energy cost for
$\alpha=1^\circ$ can be estimated by subtracting the ground state
energy from the energy of the spin tilt configuration, $J|{\bf
  s}^2|[1-\cos(1^\circ)]=1.371\mu $eV, if we choose $l=0.01\hbar$,
$J=0.1$ eV$/\hbar^2$ and $s=0.3\hbar$.  For the entire 240 sites, the
energy cost is 0.33 meV. Such a low-energy spin-wave excitation is the
origin of the strong demagnetization.  We also understand why spin
flips.  Figure \ref{fig4}(c) shows that spin-orbit torque \cite{epl16}
$\tau_z=\lambda ({\bf L}\times {\bf S})_z$ around 0 fs is positive and
large, so the spin flips from the $-z$ axis to the $+z$ axis. Figure
\ref{fig4}(d) illustrates the key idea of our alternative paradigm: It
is this spin-orbit torque that flips the spin, generates a highly
noncollinear spin wave and cancels the spins at different lattice
sites.

\section{Conclusion}

We have established an alternative paradigm for laser-induced
ultrafast demagnetization. This {\clr alternative} paradigm is based
on the spin-orbit torque-induced spin-wave excitation.  We employ a
magnetic film with over one million spins, exchange-couple them, and
take into account both spin-orbit coupling and realistic interaction
with the laser field.  By temporally and spatially resolving all the
spins, we find that spin flipping starts at the center of excitation
and generates a massive spin-wave extending across several hundred
lattice sites.  Small spin tilting from one site to next keeps the
energy cost extremely low, leading to a highly efficient
demagnetization. A pulse of 0.1 $\rm mJ/cm^2$ is capable of reducing
the spin of $0.3\hbar$ by $20\%$.  Our {\clr alternative} paradigm
differs from prior proposed mechanisms in two aspects.  First, our
model captures low-energy spin-wave excitation.  Second, we are able
to describe these spin waves of wavelength over hundred lattices.  Our
finding does not only explain the band mirroring effect
\cite{eich2017} and ultrafast magnetic phase transition within 20 fs
\cite{tengdin2018}, but opens the door to a fresh paradigm for
laser-induced ultrafast demagnetization.

\acknowledgments

 We would like to thank Drs. Z. H. Chen and L. W. Wang (U. S. Lawrence
 Berkeley National Laboratory) for sending us their preprint that
 motivated the current study. We thank Tyler Jenkins for the
 assistance with Table II.  GPZ, MM and YHB were supported by the
 U.S. Department of Energy under Contract No. DE-FG02-06ER46304 (GPZ,
 MM and YHB). Part of the work was done on Indiana State University's
 high performance Quantum and Obsidian clusters.  The research used
 resources of the National Energy Research Scientific Computing
 Center, which is supported by the Office of Science of the
 U.S. Department of Energy under Contract No. DE-AC02-05CH11231.  XSW
 acknowledges the supprt from NNSFC (No. 118742007) and the National
 Key R\&D Program of China through Grant No. 2017YFA0303202.

$^*$ guo-ping.zhang@outlook.com

\end{document}